\documentclass[preprint,superscriptaddress,20pt]{revtex4-1}

\usepackage{graphicx,natbib}
\usepackage{dcolumn}
\usepackage{bm}
\usepackage{amsmath,graphics,amssymb,psfrag}
\usepackage[colorlinks=true,linkcolor=blue,urlcolor=blue,citecolor=blue]{hyperref}
\usepackage{tikz,xcolor,hyperref}
\definecolor{lime}{HTML}{A6CE39}
\DeclareRobustCommand{\orcidicon}{%
    \begin{tikzpicture}
    \draw[lime, fill=lime] (0,0)
    circle [radius=0.16]
    node[white] {{\fontfamily{qag}\selectfont \tiny ID}};\draw[white, fill=white] (-0.0625,0.095)
    circle [radius=0.007];
    \end{tikzpicture}
    \hspace{-2mm}}
\foreach \x in {A, ..., Z}
{\expandafter\xdef\csname orcid\x\endcsname{\noexpand\href{https://orcid.org/\csname orcidauthor\x\endcsname}{\noexpand\orcidicon}}}

\begin{document}

\title{Enhanced quantum yields and efficiency in a quantum dot photocell modeled by a multi-level system }
\author{Shun-Cai Zhao\orcidA{}}
\email[Corresponding author: ]{zsczhao@126.com }
\affiliation{Department of Physics, Faculty of Science, Kunming University of Science and Technology, Kunming, 650500, PR China}

\author{Jing-Yi Chen }
\affiliation{Department of Physics, Faculty of Science, Kunming University of Science and Technology, Kunming, 650500, PR China}

\date{\today}


\begin{abstract}
To absorb the photons below the band-gap energy effectively, we proposed a quantum dot (QD) photocell modeled by multi-level system for the quantum yields and photo-to-charge efficiency limits. The theoretical results show the quantum yields are enhanced as compared to the single band-gap solar cell, and the photo-to-charge efficiencies are larger than Shockley and Queisser efficiency in the same absorbed spectrum. What's more, at the room temperature the efficiency limits are well beyond 63\(\%\) achieved by Luque and Marti (Ref\cite{26}) due to absorbing the low-energy photons via two sub-bands in this proposed photocell system. The achievements may reveal a novel theoretical approach to enhance the QD photocell performance modeled a multi-level absorbing photons system.
\begin{description}
\item[PACS numbers]42.50.Gy
\item[Keywords]Quantum dot photocell modeled by a multi-level system;  photo-to-charge efficiency; quantum yields; low-energy photons;
\end{description}
\end{abstract}
\maketitle
\section{INTRODUCTION} 
Sunlight is the most abundant energy source available on earth. Thus, solar cell attracts the enduring interests
from the scientists in the world, but the photo-to-charge efficiency limit is still a challenge for them.
Therefore, various materials and device concepts, such as the inorganic material cell\cite{1}, organic photovoltaics\cite{2,3,4,5},
hybrid perovskites\cite{6} and nanostructured solar cells\cite{7} have been proposed to enhance the conversion efficiencies
\cite{8,9}. In addition to these, one of the main research and development directions is toward reducing
the fundamental losses in the solar cells\cite{10,a}. However, a better understanding of the losses shall provide new insights into
the quantum process. Thus, quantum coherence \cite{10,11,12,13,14,15,16,17,18} has been demonstrated to break the detailed balance
and resulting in enhancement of the output power in solar cells. It has been shown that the Fano interference can lead to
suppression of absorption in an optical system\cite{12,13,14}. Recently, Fano interference reveals the enhancement of absorption light and
the power across the load\cite{15} in a photocell. A scheme of photon ratchet intermediate band solar cell\cite{b} was introduced due to the increasing lifetime of the intermediate states, which can reduce unnecessary entropy generation and the Boltzmann loss through radiative emission.
Another important direction is to find efficient absorbers for increasing the efficiency limit.
Then, split spectrum solar cells\cite{19,20}, multi-junction solar cells\cite{21} and tandem solar cells\cite{22} were introduced for the multi-photon absorption.

With the multi-photon absorption contributing to the quantum yields in heart, we propose an enhanced quantum yields scheme in a QD photocell modeled by a multi-level system, which can fully absorb the low-energy photons. And the introduced intermediate band in this proposed photocell model may be achieved by super-lattices and organized quantum dots\cite{23,24}. Thus, the separated state may be created due to quantum size effects\cite{25}, and two solar photons below the energy gap can be absorbed by two sub-bands simultaneously. As facilitates the increasing carriers forming the delocalized state in the intermediate band. Thus, the multi-photon absorption process enhances the quantum yields and the photo-to-charge efficiency limit as compared to the achieved efficiencies by Luque and Marti in 1997\cite{26}.

In the following, the work is organized as follows: the proposed quantum dot photocell model with an intermediate band is described in Sec.2. And the current-voltage characteristic , power and photo-to-charge efficiency are evaluated by different optimum parameters in Sec.3. Sec.4 presents our conclusions and outlook.

\section{The proposed quantum dot photocell model with multi-level system }

We proposed the quantum dot photocell modeled by a multi-level model generated from the theoretical prototype mentioned in Ref.\cite{13}, and an intermediate band was introduced in this multi-level quantum dot photocell. The intermediate band may be grown by super-lattices or organized quantum dots\cite{23,24}, and \(\mu_{c}\), \(\mu_{v}\) represent chemical potentials of a cathode in the conduction band and an anode in the valence band [seen Fig.1(a)]. Therefore, the single band-gap between the conduction and valence band was divided into two sub-bands, i.e., the upper sub-band gap \(E_{1I}\) and the lower sub-band gap \(E_{Ib}\). Two solar photons \(\hbar\nu_{s_{i}}\), \(\hbar\nu_{s_{j}}\) with energy below the semiconductor band gap \(E_{1b}\) are absorbed resonantly by the two subdivided energy gaps shown in Fig.1(a), which resulted in more excited electrons transfering from the intermediate band \(|I\rangle\) to the conduction band \(|c_{1}\rangle\). Fig. 1(b) is the corresponding multi-level diagram for Fig.1(a), in which the upper level \(|c_{1}\rangle\) depicts the conduction band and level \(|b\rangle\) is on behalf of the valence band, and the level  \(|I\rangle\) is interpreted as the intermediate band. When the electronic system interacts with radiation and phonon thermal reservoirs, the thermal solar photons with frequency \(\omega_{1I}\) and \(\omega_{Ib} \) are assumed to direct onto the photocell. They drive \(|c_{1}\rangle\) \(\leftrightarrow\) \(|I\rangle\) and \(|I\rangle\)\( \leftrightarrow\) \(|b\rangle\) transitions with the average occupation numbers \(n_{1}\) = \([exp( \frac{E_{1I}}{k_{B}T_{s}})-1]^{-1}\), and \(n_{2}\) = \([exp( \frac{E_{Ib}}{k_{B}T_{s}})-1]^{-1}\), where \(E_{1I}\), \(E_{Ib}\) are the two divided energy gaps and \(T_{s}\) is the solar temperature. The ambient thermal phonons at temperature \(T_{a}\) drive the two low-energy transitions \(|c\rangle\) \(\leftrightarrow\) \(|c_{1}\rangle\) and \(|v\rangle\) \(\leftrightarrow\) \(|b\rangle\). Their corresponding phonon occupation numbers are \(n_{3}\) = \([exp( \frac{E_{cc}}{k_{B}T_{a}})-1]^{-1}\) and \(n_{4}\) = \([exp( \frac{E_{vb}}{k_{B}T_{a}})-1]^{-1}\), where \(E_{cc}\) and \(E_{vb}\) are their corresponding energy gaps, respectively.

\begin{figure}[htp]
\center
\includegraphics[width=0.42\columnwidth]{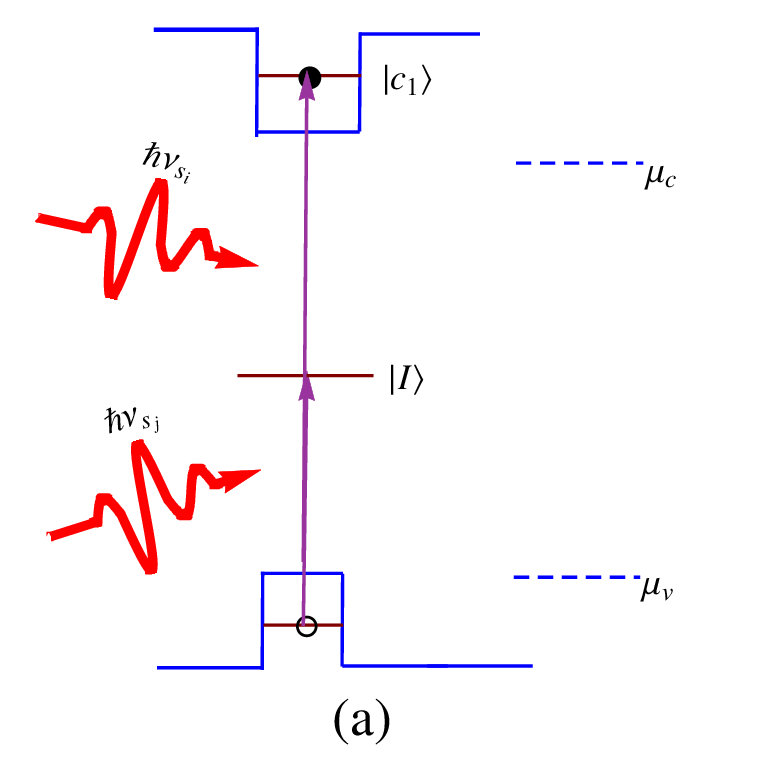 }\includegraphics[width=0.44\columnwidth]{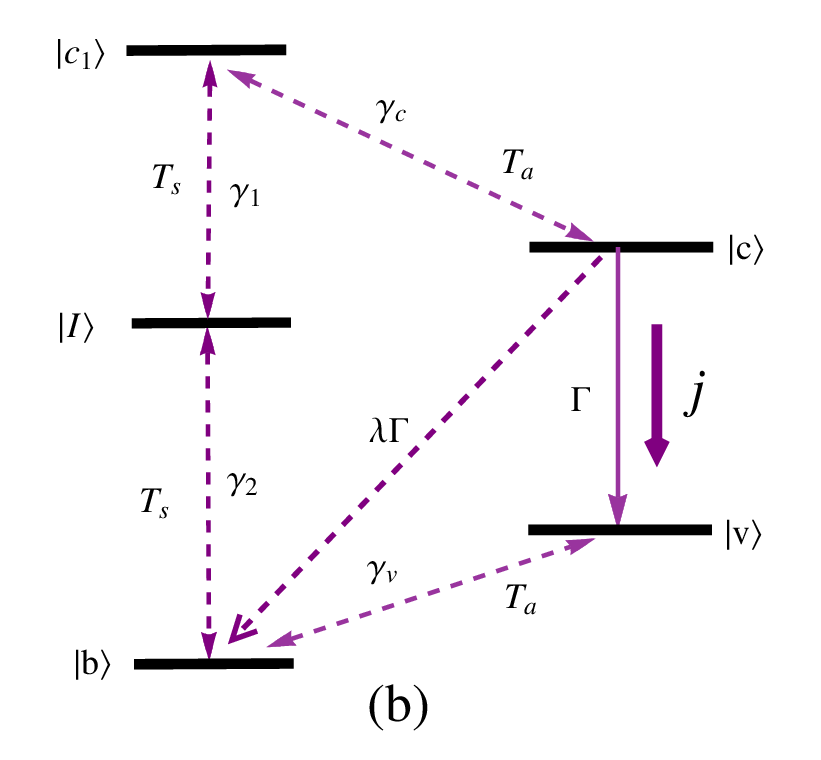 }
\caption{(Color online) (a) Quantum dots with the upper conduction band state \(|c_{1}\rangle\) and valence band state \(|b\rangle\). Two low-energy monochromatic solar photons(\(\nu_{s_{i}}\), \(\nu_{s_{j}}\)) are absorbed by two sub-band gaps divided by the intermediate band \(|I\rangle\). The host semiconductor system, in which the QDs are embedded, has effective Fermi energies \(\mu_{c}\) and \(\mu_{v}\) for the conduction and valence bands. (b) Corresponding energy levels diagram for this QD photocell model. Solar radiation drives transitions between the ground state \(|b\rangle\) and the intermediate band \(|I\rangle\), the intermediate band \(|I\rangle\) and the upper level \(|c_{1}\rangle\). Transitions \(|c_{1}\rangle\) and \(|c\rangle\), \(|v\rangle\) and \(|b\rangle\) are driven by ambient thermal phonons. Levels \(|c\rangle\) and \(v\rangle\) are connected to a load.}
\end{figure}
\label{Fig.1}

The interaction Hamiltonian describing the electronic system interacting with radiation and phonon thermal reservoirs in the rotating-wave approximation is given as follows,

\begin{eqnarray}
&\hat{V}(t)=&\hbar \{\sum_{i} g_{i}e^{i(\omega_{1I}-\nu_{i})t}\hat{\sigma}_{I1}\hat{a}_{i}+ \sum_{j} g_{j} e^{i(\omega_{Ib}-\nu_{j})t}\hat{\sigma}_{Ib} \hat{a}_{j}+ \nonumber\\
       &&     \sum_{o}g_{o}e^{i(\omega_{cc}-\nu_{o})t}\hat{\sigma}_{cc}\hat{b}_{o} +\sum_{p}g_{p} e^{i(\omega_{bv}-\nu_{p})t}\hat{\sigma}_{bv} \hat{b}_{p}\}+h.c.
\end{eqnarray}

\noindent  where \(g_{i,j,o,p}\) are the coupling constants for the corresponding transitions, \(\hbar\omega_{ij}=E_{i}-E_{j}\) is the energy spacing
between levels \(|i\rangle\) and \(|j\rangle\). \(\nu_{i}\) is the photon or phonon frequency, and \(\hat{a}_{i,j}\),  \(\hat{b}_{o, p}\) are the photon annihilation operators and phonon annihilation operators, respectively. \(\hat{\sigma}_{ij}(\hat{\sigma}_{ji})\) is the Pauli rise or fall operator. The equation of motion for the electronic density operator \(\hat{\rho}\) reads

\begin{eqnarray}
&\dot{\hat{\rho}}(t)=&-\frac{i}{\hbar} Tr_{R}[\hat{V}(t),\hat{\rho}(t_{0})\bigotimes\hat{\rho}_{R}(t_{0}]-\frac{1}{\hbar^{2}}Tr_{R} \nonumber\\
      &&    \int^{t}_{t_{0}}[\hat{V}(t),[\hat{V}(t'),\hat{\rho}(t')\bigotimes\hat{\rho}_{R}(t_{0}]]dt',
\end{eqnarray}

\noindent where the density operator \(\hat{\rho}_{R}\) describes the phonon thermal reservoirs. In the Weisskopf-Wigner approximation, the master equations
describing the the electronic system interacting with radiation and phonon thermal reservoirs are obtained in the following,
\begin{eqnarray}
&\dot{\rho_{11}}\!=\!&-\gamma_{1}[(1+n_{1})\rho_{11}-n_{1}\rho_{II}]-\gamma_{c}[(1+n_{3})\rho_{11}-n_{3}\rho_{cc}],\nonumber\\
&\dot{\rho_{II}}\!=\!&-\gamma_{2}[(1+n_{2})\rho_{II}-n_{2}\rho_{bb}]-\gamma_{1}[(1+n_{1})\rho_{11}-n_{1}\rho_{II}],\nonumber\\
&\dot{\rho_{cc}}\!=\!&\gamma_{c}[(1+n_{3})\rho_{11}-n_{3}\rho_{cc}]-\Gamma(1+\lambda)\rho_{cc},\\
&\dot{\rho_{vv}}\!=\!&\Gamma\rho_{cc}+\gamma_{v}n_{4}\rho_{bb}-\gamma_{v}(1+n_{4})\rho_{vv}, \nonumber
\end{eqnarray}

\noindent where \(\gamma_{i}\) is modeled the electrons' decaying process as spontaneous decay rates of the corresponding transitions [see Fig.1(b)]. In this model, levels \(|c\rangle\) and \(|v\rangle\) are connected to a load, and the load is yielding a decay of the level \(|c\rangle\) into level \(|v\rangle\) at a rate \(\Gamma\). The recombination between the acceptor and the donor is modeled by a decay rate of \(\lambda \Gamma\) [also see Fig. 1(b)], where \(\lambda\) is a dimensionless fraction. And we focus on steady-state operation of the photocell model, voltage \(V\) across the load is expressed in terms of populations of the levels \(|c\rangle\) and \(|v\rangle\) as\cite{11,12,13,14},

\begin{eqnarray}
eV\!=\!  E_{cv}+k_{B}T_{a} \ln(\frac{\rho_{cc}}{\rho_{vv}}).
\end{eqnarray}

\noindent where \(E_{cv}=E_{c}-E_{v}\), \(T_{a}\) is the ambient temperature circumstance. The current \(j\) through the cell is interpreted as  \(j=e \Gamma\rho_{cc}\) and power delivered to the
load is \(P\) = \(j ¡¤ V \) which is supplied by the incident solar radiation power \(P_{s}\)\cite{11}. Then, the the photo-to-charge efficiency can be calculated as the ratio between the output power \(P\) and incident solar radiation power \(P_{s}\),
\begin{eqnarray}
\eta \!=\! \frac{P}{P_{s}}=\frac{V \cdot j(V,T_{s},T_{a},E_{g})}{P_{s}} .
\end{eqnarray}
\noindent where \(E_{g}\) describes the different energy gaps, and the incident solar radiation power \(P_{s}\) equals to \(\frac{\hbar(\nu_{i}+\nu_{j})}{e}\) with the fundamental charge of an electron being \(-e\).

\section{ Quantum yields within the energy gap range [0.5\(ev\), 1.25\(ev\)]}

Shockley and Queisser efficiency\cite{27} indicates that semiconductors can present the photovoltaic effect within band gaps ranging from about 1.25 \(eV\) to 1.45 \(ev\), while the solar spectrum in the range of [354\(nm\), 2480\(nm\)] contains photons with energies ranging from 0.5 ev to 3.5 ev. Photons with energy below the semiconductor band gap, i.e., in the range of [0.5\(ev\), 1.25\(ev\)] are not absorbed. In this work, we discuss the quantum yields with two sub-band energy gaps falling in the range of [0.5\(ev\), 1.25\(ev\)], in which the solar photons are called low-energy photons. The radiative recombination between the acceptor and the donor is often regarded as the fundamental losses in the cells, and it intuitively affects linearly the quantum yields. Fig.2(a) shows our results about the quantum yields versus the voltage with the optimum values of \(\lambda\) at the room temperature \(T_{a}\)=0.0259\(ev\), and their corresponding current-voltage characteristic diagram is inserted in Fig.2(a). The lengths of black double-arrow lines in the insert diagram indicate that the radiative recombination decreases the current-voltage characteristic nonlinearly. The nonlinear reduction can also be shown by the peak power curves in Fig.2(a). These counter-intuitive results demonstrate that there is other quantum loss in the quantum yields except the radiative recombination. We also notice that the output current is about 0.038 and the corresponding peak power is about 0.048 at \(\lambda=0.75\), which are larger than those with full noise-induced coherence in Ref \cite{13}, respectively. These results indicate two sub-bands can substantially enhance the quantum yields. The underlying mechanism is that the carriers delocalized in the intermediate sate \(|I\rangle\) can increase the possibility of carriers transiting to the conduction band, which ultimately improves the quantum yields greatly.

\begin{figure}[!t]
\centerline{\includegraphics[width=0.48\columnwidth]{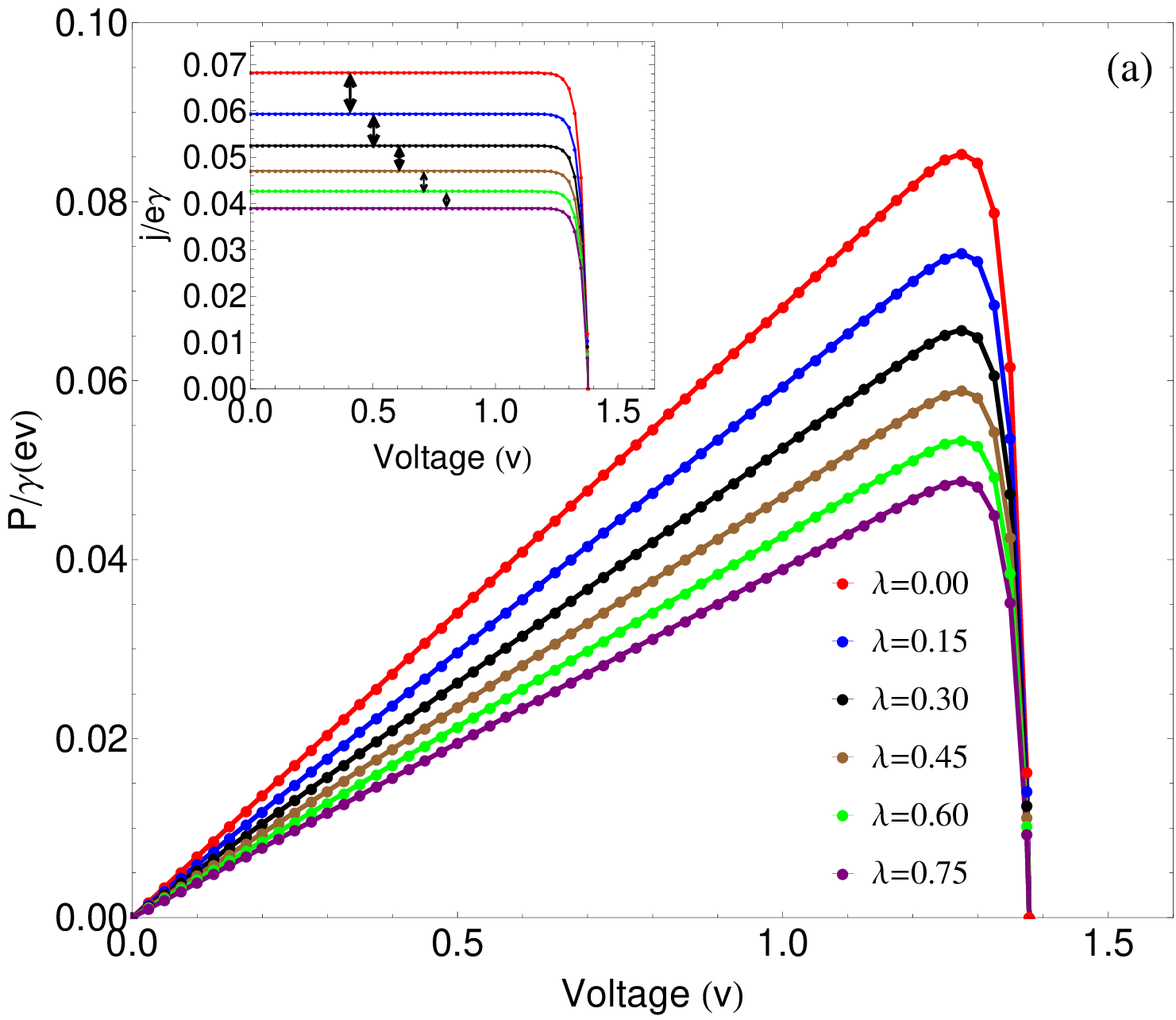 }\includegraphics[width=0.48\columnwidth]{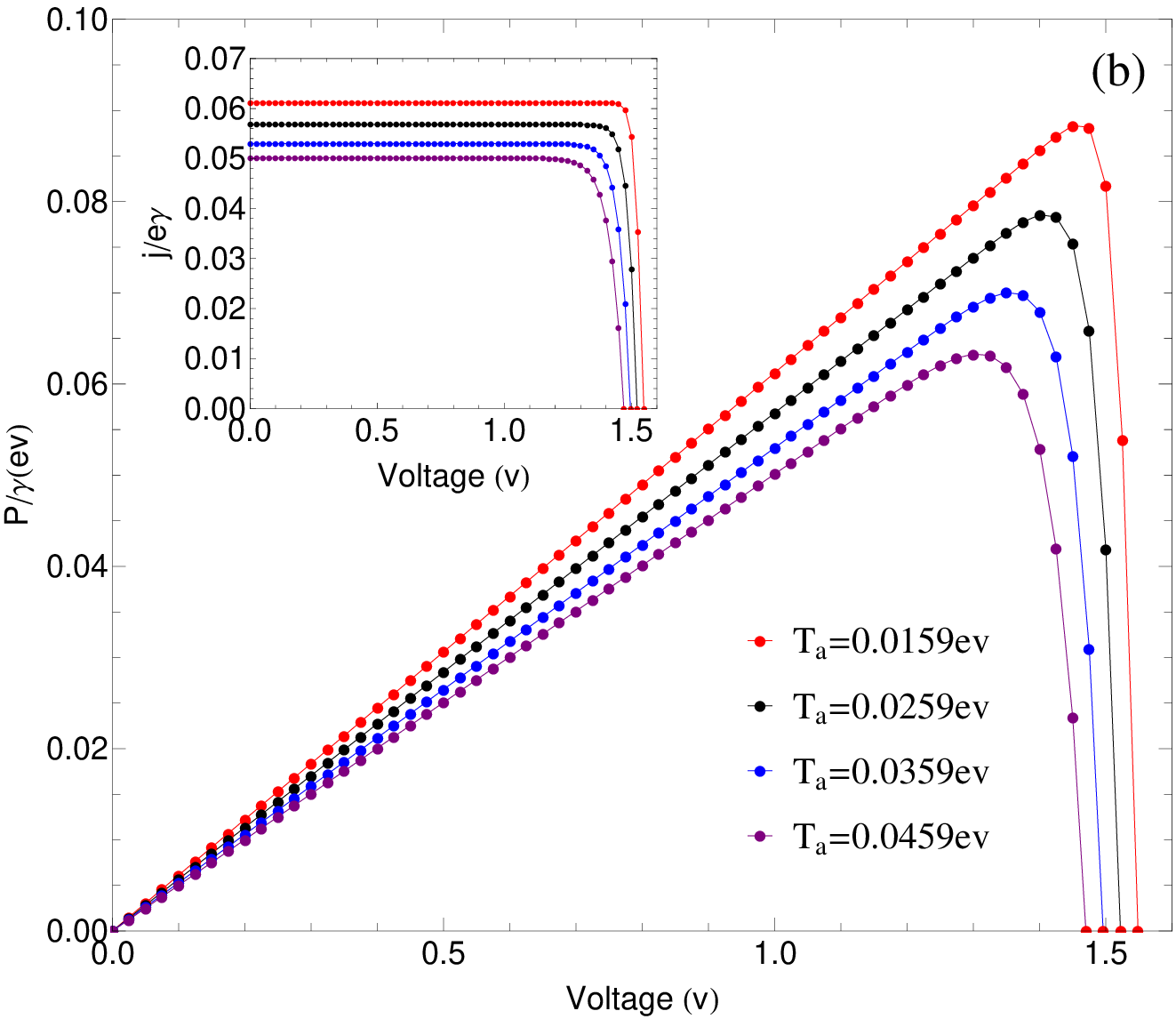}}
 \caption{(Color online) Current-voltage characteristic and power generated by the QD photocell modeled by a multi-level system versus the induced cell voltage V for (a) different radiative recombination rates \(\lambda\) and (b) for different ambient temperatures \(T_{a}\) with \(\lambda=0.15\). It takes \(E_{cv}\)=1.43\(ev\), \(E_{1I}\)=0.55\(ev\), \(E_{Ib}\)=0.60\(ev\), \(E_{cc}\)=\(E_{vb}\)=0.005\(ev\), \(T_{s}\)=0.5\(ev\), \(\gamma_{2}\)=10\(\gamma_{1}\),\(\gamma_{c}\)=\(\gamma_{v}\)=50\(\gamma_{1}\),\(\gamma_{1}\) is the scale unit.}
\label{fig2}
\end{figure}

The ambient circumstance temperature selected as the optimum parameters is shown in Fig.2(b). And one can see that some different features are manifested by the curves of the current-voltage characteristic and output power. The output current and peak power decrease with the ambient temperatures \(T_{a}\), and the data displays that the current declines from 0.062 to 0.05 and the peak power varies from 0.085 to 0.063 when the ambient circumstance temperature increases by 0.01 \(ev\). So, the higher ambient circumstance temperature decreases the quantum yields prominently. In this proposed photocell model, we also notice that the quantum yields, i.e., current-voltage characteristic and output power are significantly improved as compared to those in Ref \cite{13} at \(T_{a}\)=0.0259\(ev\), and the peak power is more than twice as much as that in Ref\cite{13} with the full coherence. What's more, the higher ambient circumstance temperature causes the shrinks of the voltages in Fig.2(b), which is different from the radiative recombination effects in Fig.2(a). These results demonstrate that sub-bands absorbed photons within the low-energy region can enhance the quantum yields greatly. However, the quantum behavior of the charge carriers (electrons and holes) in the transporting to the conduction band is greatly affected by the ambient temperature.

\begin{figure}
\centerline{\includegraphics[width=0.48\columnwidth]{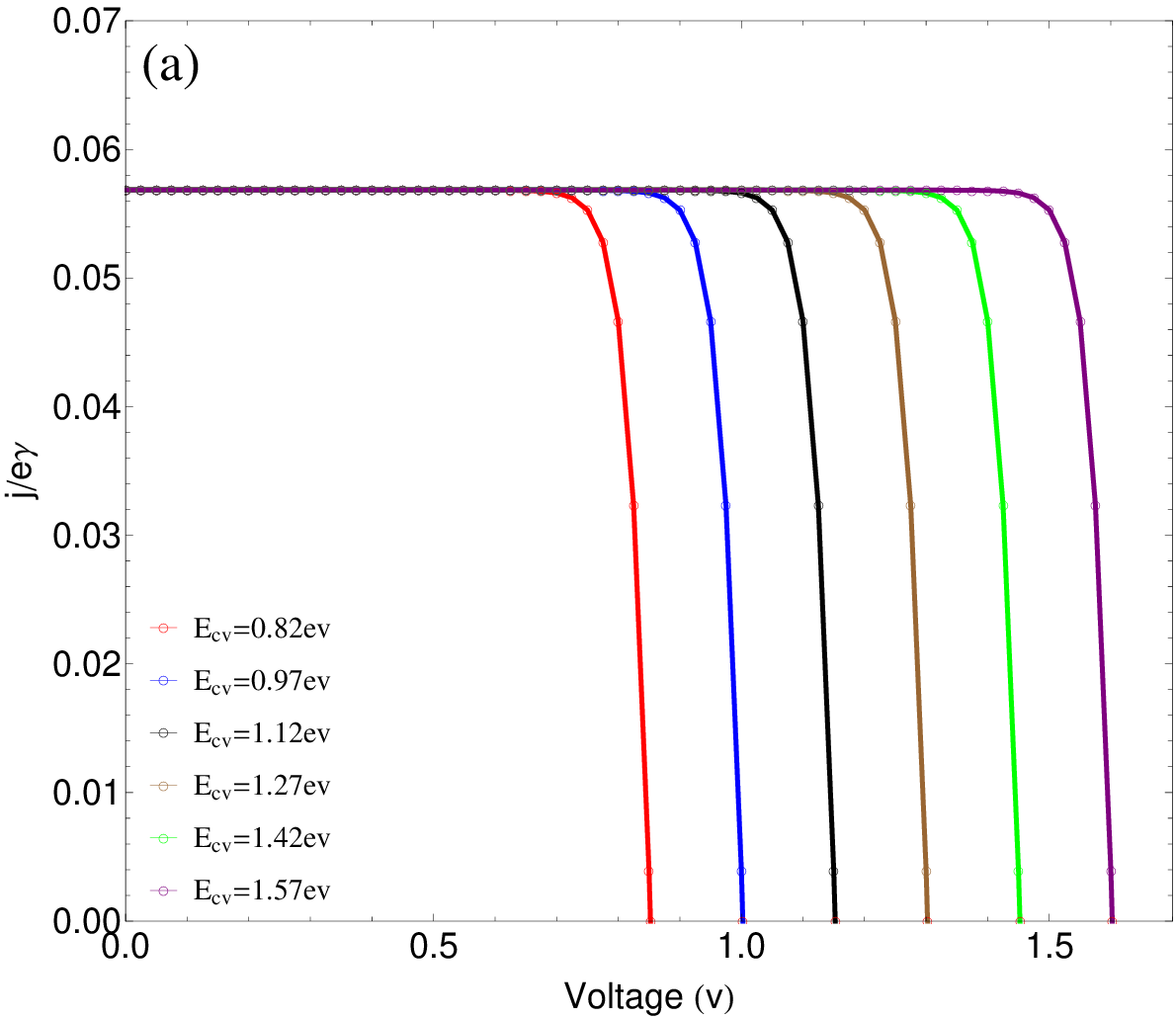}\includegraphics[width=0.48\columnwidth]{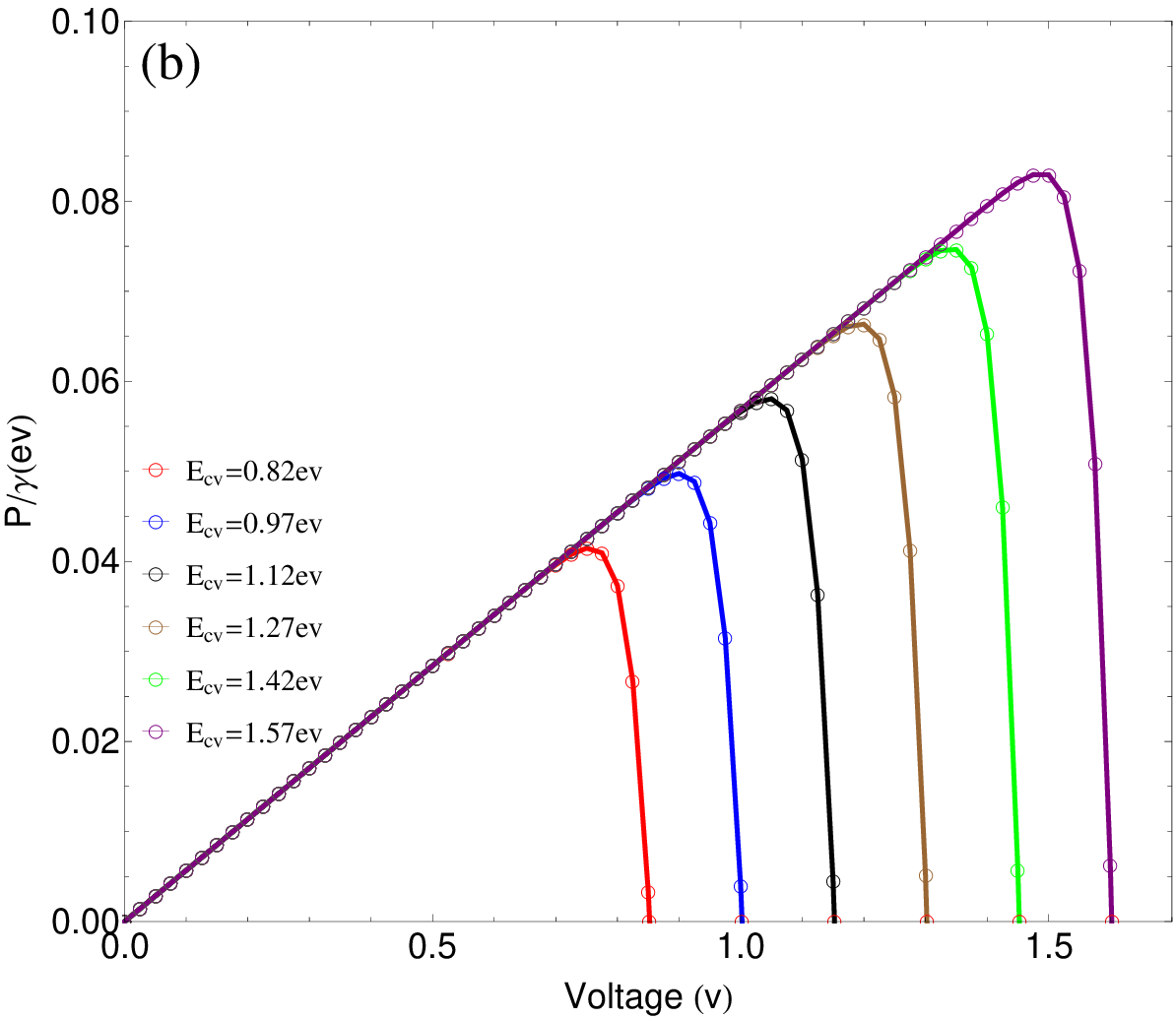}}%
\caption{(Color online) (a) Current-voltage characteristic (b) power generated by the QD photocell modeled by a multi-level system as a function of induced cell voltage \(V\) with different \(E_{cv}\). \(E_{1I}\)=0.55\(ev\),\(E_{Ib}\)=0.75\(ev\), \(T_{a}\)=0.0259\(ev\), \(\lambda=0.35\), other parameters are the same to those in Fig.2.}
\label{Fig.3}
\end{figure}

We consider the influence of energy spacing \(E_{cv}\) on the quantum yields in Fig.3. The other selected parameters are the same as those in Fig.2. It shows that the currents remain the same values while the voltage increases in tandem with the optimum value of \(E_{cv}\) by the current-voltage characteristic curves (Seen in Fig.3(a)). The peak power and voltage monotonically increase with the optimum values of \(E_{cv}\). The optimum energy spacing \(E_{cv}\) displays a linear and monotonic enhance for the voltage and for the peak power, while it sprinkles no effect on the output current. When the optimum value of \(E_{cv}\) is large enough the photocell generates more power. This provides an approach to improve the output voltage and peak power while outputs the steady current.

The introduced intermediate band is the characteristic of this QD photocell with multi-level system, and it provides the optimum values of the two sub-band gaps. The previous results demonstrate the sub-band gaps absorbing the low-energy solar photons can enhance the quantum yields greatly. Now the question, that whether the optimum values of the upper or lower sub-band energy gaps affect the quantum yields identically or not comes out, and Fig.4 and Fig.5 give the answers with the synchronous optimization of the upper and lower sub-band energy gaps, \(E_{1I}\) and \(E_{Ib}\). Although the current-voltage characteristic displays the same evolutionary characteristics versus the voltage in Fig.4(a) and Fig.5(a), i.e., the output currents decrease with the upper or lower sub-band gaps, \(E_{1I}\) or \(E_{Ib}\), we also notice that the output currents in Fig.4(a) are larger than those in Fig.5(a) with the same optimum values of the two sub-band energy gaps. When \(E_{1I}\)=\(E_{Ib}\)=0.5\(ev\), the output current is 0.085\(e\gamma\) in Fig.4(a) and 0.066\(e\gamma\) in Fig.5(a). It indicates that the different function between the lower and upper sub-band gaps and that the low-energy solar photons absorbed by the lower sub-band contribute the current more than those by the upper sub-band. The comparison between Fig.4(a) and Fig.5(a) reveals the underlying mechanism that the narrower lower sub-band can absorb much more low-energy solar photons shored in the intermediate band, which promotes the possibility for carriers transiting to the conduction band and forming the output current.

\begin{figure}
\centerline{\includegraphics[width=0.48\columnwidth]{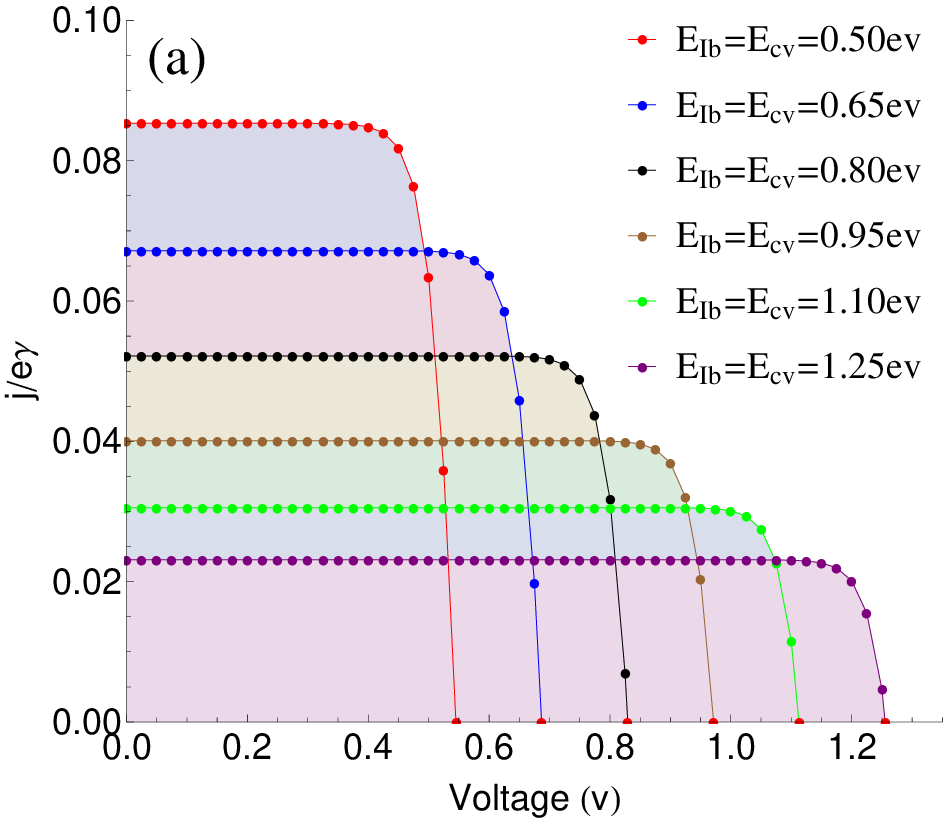}\includegraphics[width=0.48\columnwidth]{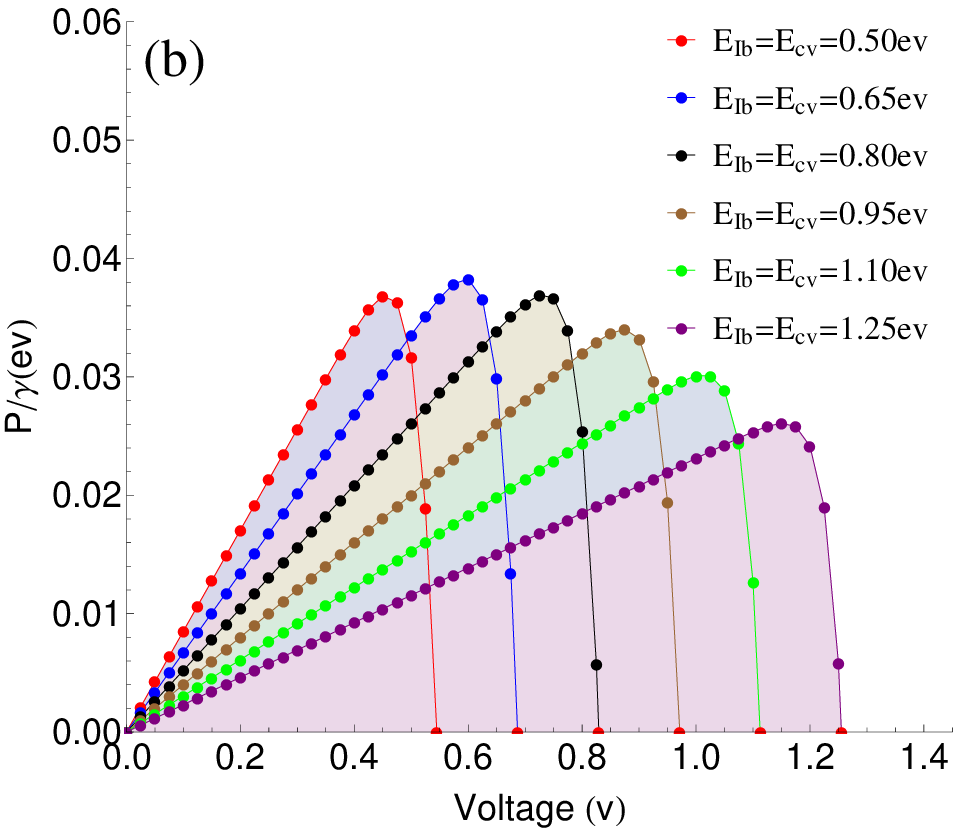}}%
\caption{(Color online) (a) Current-voltage characteristic and (b) power generated by the QD photocell modeled by a multi-level system versus the output voltage \(V\) under the different lower sub-band energy gaps \(E_{Ib}\). Other parameters are the same to those in Fig.3.}
\label{Fig.4}
\end{figure}

\begin{figure}
\centerline{\includegraphics[width=0.48\columnwidth]{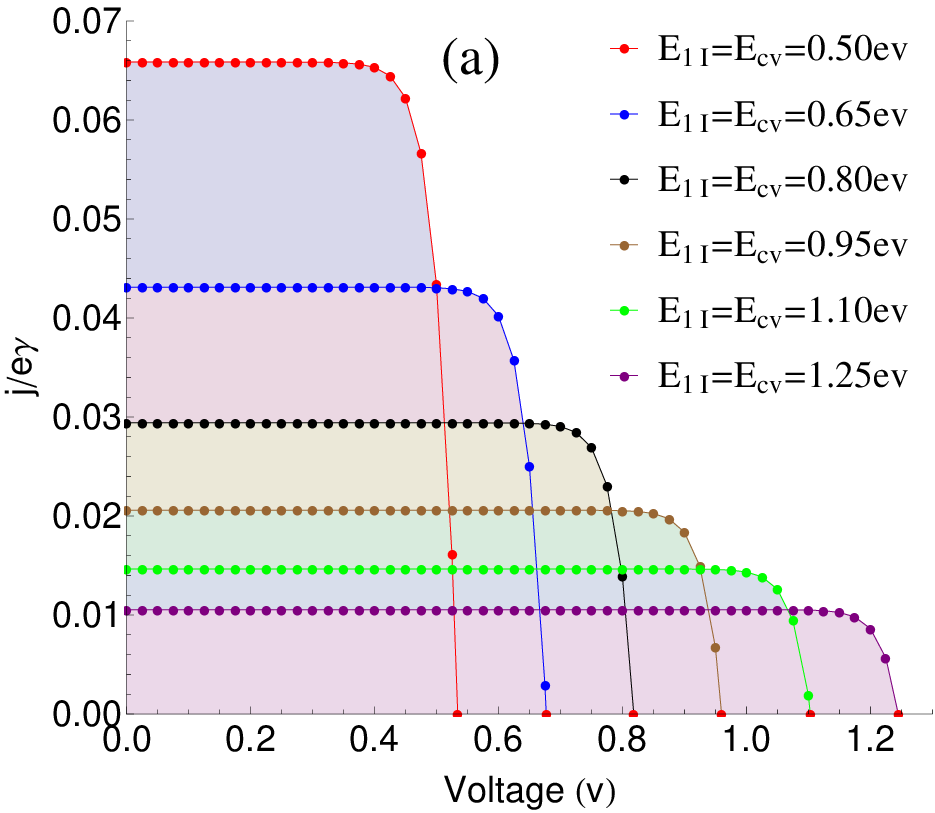}\includegraphics[width=0.48\columnwidth]{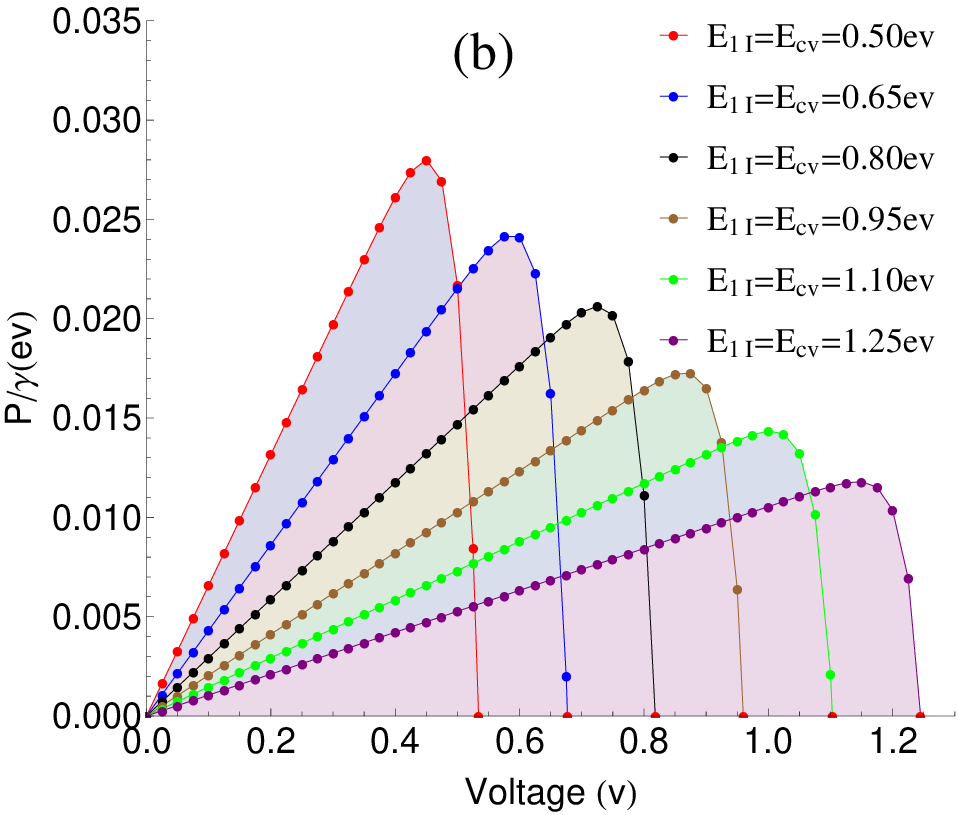}}%
\caption{(Color online) (a) Current-voltage characteristic and (b) power generated by the QD photocell modeled by a multi-level system versus the output voltage \(V\) under the different upper sub-band energy gaps \(E_{1I}\). Other parameters are the same to those in Fig.3.}
\label{Fig.5}
\end{figure}

However, the influences between the lower and upper sub-bands \(E_{Ib}\), \(E_{1I}\) are not completely the same on the output peak power in Fig.4(b) and Fig.5(b). The output peak power monotonously decreases with the upper sub-band \(E_{1I}\) in Fig.5(b), while there is a slight increment then decrease in the output peak power during the increment of the lower sub-band \(E_{Ib}\) in Fig.4(b). These results signify much fewer low-energy solar photons are absorbed by the wider upper sub-band \(E_{1I}\) gap in the energy gap of [0.5\(ev\), 1.25\(ev\)], and that much fewer carriers produced in the conduction band incurs the decreasing output peak power. The wider band-gap of the lower sub-band can absorb much high-energy solar photons in the range [0.5\(ev\), 1.25\(ev\)], but the much higher-energy solar photons absorbed by the wider lower sub-band \(E_{Ib}\) lose as multiple phonon emission because of the photo-generated carriers with much larger kinetic energy. These cause to the slight increment then decrease in the peak power in Fig.4(b).

\section{Photo-to-charge efficiency within solar spectrum }

As is well known, semiconductors generate the Shockley and Queisser efficiency\cite{27} within the band-gap ranging from about 1.25 \(ev\) to 1.45 \(ev\). That's to say, the absorbed spectrums for the Shockley and Queisser efficiency ranges about in [855 \(nm\), 992 \(nm\)] (the green areas shown in Fig.6 (a) and (b)). However, the solar spectrum ranges from about 354 \(nm\) to 2480 \(nm\). In Fig.6, we simulate the two sub-bands \(E_{1I}\) and \(E_{Ib}\) to absorb all the solar photons from the solar spectrum, and examine the photo-to-charge efficiency \(\eta\) with different optimum ambient temperatures \(T_{a}\).

\begin{figure}
\centerline{\includegraphics[width=0.48\columnwidth]{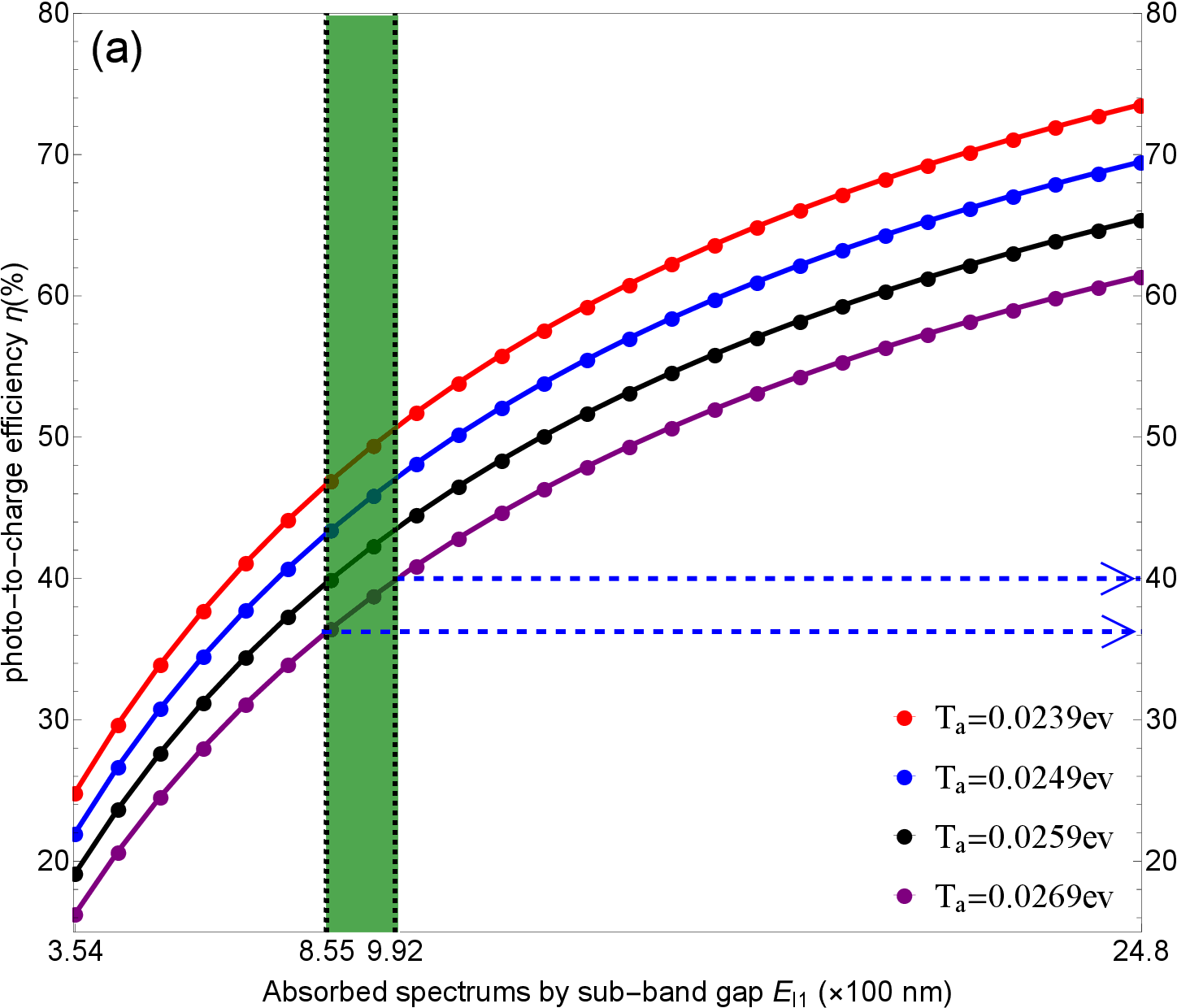}\includegraphics[width=0.48\columnwidth]{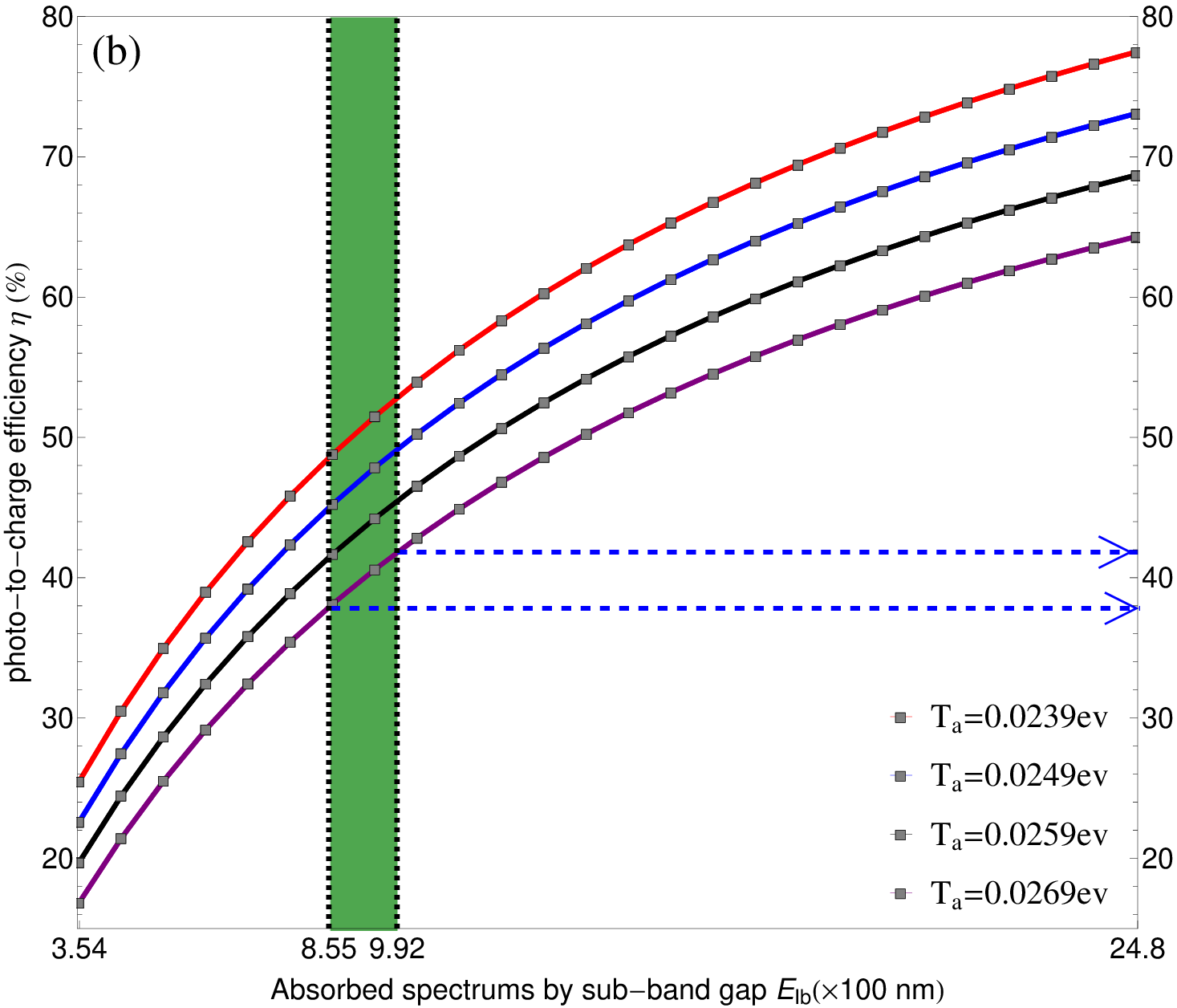}}%
\caption{(Color online) Photo-to-charge efficiency \(\eta\) generated by the QD photocell modeled by a multi-level system versus the absorbed spectrums by (a) the upper sub-band \(E_{1I}\)  and by (b) the lower sub-band \(E_{Ib}\) with the optimum values of the ambient temperatures \(T_{a}\). It takes \(E_{Ib}\)=1.35\(ev\) in (a) and \(E_{1I}\)=1.25 \(ev\) in (b),  \(E_{cv}\)=1.49\(ev\), other parameters are the same to those in Fig.3.}
\label{Fig.6}
\end{figure}

Investigation to Fig.6 (a) and (b), three significant results were achieved straightly. Firstly, the photo-to-charge efficiency \(\eta\) decreases with the ambient temperatures \(T_{a}\). What's more, the influence on \(\eta\) caused by \(T_{a}\) is much stronger in the long wavelength region than those in the short wavelength region, which can be drawn from the spacings between the curves both in Fig.6 (a) and (b). As demonstrates that the higher energy photons in the short wavelength region have the greater possibility to transit to the conduction band and bring out little efficiency variation in a higher circumstance temperature. The larger \(\eta\) in the long wavelength region indicates the absorbed low-energy photons is a good optimum approach to enhance \(\eta\) in this proposed QD photocell model. Secondly, \(\eta\) in this proposed QD photocell are larger than the Shockley and Queisser efficiency in green areas with photons energy ranging in [1.25\(ev\), 1.45\(ev\)]. At \(T_{a}\)=0.0269\(ev\) in the green areas shown by Fig.6 (a) and (b), this proposed QD photocell outputs the smallest \(\eta\) with the different optimum values of \(T_{a}\). The values of \(\eta\) are about 36\(\%\) and 38\(\%\) with absorbing wavelength 855 \(nm\), and 40\(\%\) and 42\(\%\) with absorbing wavelength 992 \(nm\) in Fig.6 (a) and (b), respectively. The result owns to the carriers generated by the two sub-band gaps absorbing more low-energy photons than those in the Shockley and Queisser efficiency model. We also achieve the third knowledge from the comparison of the absorbed spectrums by \(E_{1I}\) with \(E_{Ib}\). At the ambient circumstance temperature \(T_{a}\)=0.0269\(ev\), the values of \(\eta\) at the wavelength 855 \(nm\) and 992 \(nm\) in Fig.6 (a) are smaller than those in Fig.6(b), respectively. Which shows the different roles of the two sub-bands \(E_{1I}\) and  \(E_{Ib}\) in the output efficiency in this proposed photocell model. Not only that, but the photo-to-charge efficiency limits are about 65.42\(\%\) and 68.88\(\%\) at the room temperature \(T_{a}\)=0.0259 \(ev\) in  Fig.6 (a) and (b), respectively, which is larger than the efficiency limit 63\(\%\) achieved by Luque and Marti in 1997\cite{26}.

Before concluding this paper, we would like to emphasize the following points. First, our current work is focused on how the sub-band energy gaps to affect the quantum yields. Furthermore, our central results indicate that the enhanced quantum yields are achieved as compared to the Ref\cite{13} by the absorbed low-energy photons, and the photo-to-charge efficiency \(\eta\) is larger than the Shockley and Queisser efficiency with different optimum ambient temperatures. Second, we only discuss the photons absorbed resonantly by this proposed QD photocell, while for energies above the band gap charge carriers with an excess kinetic energy are not considered here. Third, the interband transition rate of the carriers is also not discussed here. But the above mentioned points does not influence to reveal the significance of the current work. And we believe our future work will bring some important insights into these fundamental problems.

\section{Summary and discussion }

In summary, by simulating a QD solar cell with a multi-level system, we discussed its quantum yields with the sub-bands absorbing solar photons ranging in [0.5\(ev\), 1.25\(ev\)].
Under different optimum parameters, the quantum yields were enhanced as compared to the single band-gap quantum dot photocell. In the same solar absorbed spectrum region, the achieved photo-to-charge efficiencies are larger than the Shockley and Queisser efficiency, and the efficiency limits at the room temperature are also larger than 63\(\%\) obtained by Luque and Marti. Thus, it points out a novel approach to enhance the QD photocell performance via the full absorbing the low-energy photons.

\begin{acknowledgments}
We thank the financial supports from the National Natural Science Foundation of China ( Grant Nos. 61205205 and 61565008 ), and
the General Program of Yunnan Applied Basic Research Project, China ( Grant No. 2016FB009 ).
\end{acknowledgments}






\bibliographystyle{alpha}

\end{document}